\documentclass[epsfig,12pt]{article}
\usepackage{epsfig}
\usepackage{graphicx}
\usepackage{hyperref}

\usepackage{array}
\usepackage{amsmath}
\usepackage{amssymb}

\newcommand{\beq}{\begin{equation}}   
\newcommand{\eeq}{\end{equation}}
\newcommand{\beqn}{\begin{eqnarray}}   
\newcommand{\eeqn}{\end{eqnarray}}

\begin{document}
\unitlength = 1mm

\def\de{\partial}
\def\Tr{ \hbox{\rm Tr}}
\def\const{\hbox {\rm const.}}  
\def\o{\over}
\def\im{\hbox{\rm Im}}
\def\re{\hbox{\rm Re}}
\def\bra{\langle}\def\ket{\rangle}
\def\Arg{\hbox {\rm Arg}}
\def\Re{\hbox {\rm Re}}
\def\Im{\hbox {\rm Im}}
\def\diag{\hbox{\rm diag}}


\def\QATOPD#1#2#3#4{{#3 \atopwithdelims#1#2 #4}}
\def\stackunder#1#2{\mathrel{\mathop{#2}\limits_{#1}}}
\def\stackreb#1#2{\mathrel{\mathop{#2}\limits_{#1}}}
\def\Tr{{\rm Tr}}
\def\res{{\rm res}}
\def\Bf#1{\mbox{\boldmath $#1$}}
\def\balpha{{\Bf\alpha}}
\def\bbeta{{\Bf\beta}}
\def\bgamma{{\Bf\gamma}}
\def\bnu{{\Bf\nu}}
\def\bmu{{\Bf\mu}}
\def\bphi{{\Bf\phi}}
\def\bPhi{{\Bf\Phi}}
\def\bomega{{\Bf\omega}}
\def\blambda{{\Bf\lambda}}
\def\brho{{\Bf\rho}}
\def\bsigma{{\bfit\sigma}}
\def\bxi{{\Bf\xi}}
\def\bbeta{{\Bf\eta}}
\def\d{\partial}
\def\der#1#2{\frac{\d{#1}}{\d{#2}}}
\def\Im{{\rm Im}}
\def\Re{{\rm Re}}
\def\rank{{\rm rank}}
\def\diag{{\rm diag}}
\def\2{{1\over 2}}
\def\ntwo{${\mathcal N}=2\;$}
\def\nfour{${\mathcal N}=4\;$}
\def\none{${\mathcal N}=1\;$}
\def\ntwot{${\mathcal N}=(2,2)\;$}
\def\ntwoo{${\mathcal N}=(0,2)\;$}
\def\x{\stackrel{\otimes}{,}}

\newcommand{\cpn}{CP$(N-1)\;$}
\newcommand{\wcpn}{wCP$_{N,\widetilde{N}}(N_f-1)\;$}
\newcommand{\wcpd}{wCP$_{\widetilde{N},N}(N_f-1)\;$}
\newcommand{\wcpt}{$\mathbb{WCP}(2,2)\;$}
\newcommand{\wcpo}{$\mathbb{WCP}(1,1)\;$}
\newcommand{\wcp}{$\mathbb{WCP}(N,\tilde N)\;$}
\newcommand{\vp}{\varphi}
\newcommand{\pt}{\partial}
\newcommand{\tN}{\widetilde{N}}
\newcommand{\ve}{\varepsilon}
\renewcommand{\theequation}{\thesection.\arabic{equation}}

\newcommand{\sun}{SU$(N)\;$}

\setcounter{footnote}0

\vfill

\begin{titlepage}

\begin{flushright}
\end{flushright}

\begin{center}
{  \Large \bf  
 NS Three-form Flux Deformation for the 
\\[2mm]
Critical Non-Abelian Vortex String
 }

\vspace{5mm}

{\large  \bf A.~Yung$^{\,\,a,\,b}$}
\end {center}

\begin{center}

$^{a}${\it National Research Center ``Kurchatov Institute'', 
Petersburg Nuclear Physics Institute, Gatchina, St. Petersburg
188300, Russia}\\
$^{b}$ {\it Higher School of Economics, National Research University, St. Petersburg 194100, Russia}\\

\end{center}

\vspace{1cm}

\begin{center}
{\large\bf Abstract}
\end{center}

It has been  shown that  non-Abelian solitonic vortex string supported in four-dimensional (4D)
\ntwo supersymmetric QCD (SQCD) with the U(2) gauge group
and $N_f=4$  quark flavors  becomes a critical superstring. This string propagates in the ten-dimensional space formed by a product of the flat 4D space and an internal space given by a Calabi-Yau noncompact threefold, namely, the conifold.
The spectrum of low lying closed string states in the associated type IIA string theory was found and interpreted
as a spectrum of hadrons in 4D \ntwo  SQCD. In particular,
the lowest string state appears to be a massless BPS baryon associated with the deformation of the complex structure modulus
$b$ of the conifold. In the previous work the deformation of the 10-dimensional background with nonzero Neveu-Schwarz 3-form flux was considered and interpreted as a switching on a particular choice of quark masses in 4D SQCD. This deformation was studied  to the leading order at small 3-form flux. In this paper we study the back reaction of the nonzero 
3-form flux on the metric and the dilaton  introducing ansatz with several warp factors and  solving gravity equations of motion. We show that 3-form flux produces a potential for the  conifold complex structure modulus  $b$, which leads to the  runaway vacuum. At the runaway vacuum warp factors disappear, while the conifold degenerates. In 4D SQCD we relate this to the flow to the U(1) gauge theory upon switching on quark masses and decoupling of two flavors.

\vspace{2cm}

\end{titlepage}

\newpage


\newpage

\section {Introduction }
\label{intro}
\setcounter{equation}{0}

Non-Abelian vortices were first found in 
4D \ntwo  SQCD with the gauge group U$(N)$ and $N_f \ge N$ flavors of quarks
\cite{HT1,ABEKY,SYmon,HT2}. The non-Abelian vortex string is 1/2
Bogomolny-Prasad-Sommerfeld (BPS) saturated and, therefore,  has \ntwot supersymmetry on its world sheet.
In addition to four translational moduli  of the  Abrikosov-Nielsen-Olesen (ANO) strings 
\cite{ANO}, the non-Abelian string carries orientational  moduli, as well as the size moduli if $N_f>N$
\cite{HT1,ABEKY,SYmon,HT2} (see \cite{Trev,Jrev,SYrev,Trev2} for reviews). 

It was shown in \cite{SYcstring} that the non-Abelian solitonic vortex string in  
  \ntwo supersymmetric QCD (SQCD) with the U(N=2) gauge group and  $N_f=4$   flavors of quark hypermultiplets becomes a critical  superstring. The dynamics of the internal orientational and size
moduli of the non-Abelian vortex string for  the case $N =2$, $N_f=4$ is described by the so-called two-dimensional
(2D) weighted CP sigma model, which we denote as $\mathbb{WCP}(N=2,N_f-N=2)$. 

For $N_f=2N$
this world sheet sigma model becomes conformal. Moreover, for $N=2$ the 
number of the orientational and size moduli  is six and they can be combined with 
four translational moduli to form a ten-dimensional (10D) space required for a
superstring to become critical \cite{SYcstring,KSYconifold}. In this case the target space of the world sheet 
 sigma model on 
the non-Abelian vortex string is
 $\mathbb{R}^4\times Y_6$, where $Y_6$ is a non-compact six dimensional Calabi-Yau (CY) manifold, the conifold 
\cite{Candel,NVafa}. Moreover, the theory of the critical vortex string at hand was identified as the superstring theory of type IIA \cite{KSYconifold}. This allows one to apply the string theory  for the calculation  of the spectrum of  string states  and identify it with a spectrum of hadrons in 4D \ntwo SQCD \cite{KSYconifold}. Since Non-Abelian vortex strings are topologically stable and cannot be broken (see \cite{SYrev} for a review) we focus on the 
closed strings and consider Kaluza-Klein reduction of 10D string theory associated with the non-Abelian vortex to 4D.

A version of the string-gauge duality for 4D SQCD was proposed \cite{SYcstring}: at
weak coupling this theory is in the Higgs phase and can be described in terms
of quarks and Higgsed gauge bosons, while at strong coupling hadrons of this theory can be understood as closed string states formed by the non-Abelian vortex string. We  call this approach ''solitonic string-gauge duality''.

The first step of the above  program, namely,  finding massless
string states was carried out in \cite{KSYconifold,KSYcstring} using supergravity approximation.
It turns out that  most of massless modes have  non-normalizable wave functions over the non-compact conifold $Y_6$, i.e. they are not localized in 4D 
and, hence, cannot be interpreted as dynamical states in 4D SQCD. In particular, the 4D graviton and  unwanted vector multiplet associated with deformations of the K\"ahler form of the conifold are absent.
 However, a single massless BPS hypermultiplet   was found 
at the  self-dual point at strong coupling. It is associated with deformations of a complex structure of the conifold and was  interpreted  as a composite 4D baryon $b$ \footnote{ The definition of the baryonic charge is non-standard and will be given below in Sec. \ref{sec:NAstring}.}.

Later  low lying massive non-BPS 4D states were found in \cite{SYlittles,SYlittmult} using the little string theory 
approach, see \cite{Kutasov} for a review.

In the previous work \cite{Y_NSflux} a study of possible flux deformations of the 10D background for non-Abelian vortex string was initiated. The goal is to look for  flux deformations of the string background which do not destroy \ntwo supersymmetry in 4D and interpret them in terms of  certain deformations in SQCD. Fluxes generically induce a potential for CY moduli lifting flat directions, see, for example, \cite{Louis} for a review. It is known that for type IIA CY compactifications the potential for the  K\"ahler form moduli arise from Ramond-Ramond (RR) even-form fluxes,
while  the potential for complex structure moduli  is induced by the Neveu-Schwarz (NS) 3-form flux $H_3$  \cite{Louis2,Kachru}.
Since for the conifold case at hand the only modulus associated with a physical state is the complex structure modulus $b$ we focus on the NS 3-form flux. It does not break \ntwo supersymmetry in 4D theory \cite{Louis2}.

In \cite{Y_NSflux} the NS 3-form flux $H_3$ was interpreted as switching on quark masses in 4D SQCD.
The reason is that the only scalar potential deformation, which is allowed in SQCD by \ntwo supersymmetry 
is the mass term for quarks. Field theory arguments were used to find a particular choice of nonzero quark masses associated with $H_3$.

The flux deformation was studied  in \cite{Y_NSflux} to the leading order at small $H_3$ which translates into small values of quark masses. In this paper we study the back reaction of the nonzero 
3-form flux on the metric and dilaton. We introduce ansatz with several warp factors and  solve gravity equations of motion
for arbitrary value of $H_3$.
This allows us to switch on large masses for certain flavors in 4D SQCD and consider the decoupling limit.

 We show that 3-form flux produces a potential for the  conifold complex structure modulus  $b$, which leads to the  runaway vacuum. At the runaway vacuum warp factors disappear, while the deformed conifold degenerates. In 4D SQCD we relate this to the flow to U(1) gauge theory upon switching on quark masses and decoupling of two flavors. 

Note that we  assume that the conifold complex structure modulus  $b$ is large enough to make sure that
the curvature of the conifold is everywhere small. This justify the gravity
approximation.

The paper is organized follows. In Sec. \ref{sec:NAstring} we briefly review 4D \ntwo SQCD and  the world sheet sigma  model
on the non-Abelian string. Next we review massless baryon $b$ as a deformation of the complex structure of the conifold.  In Sec. \ref{sec:gravity_eqs}  we introduce the metric ansatz and solve gravity
equations of motion with nonzero 3-form $H_3$ in the limit of large radial coordinate of the conifold.  In 
Sec. \ref{sec:large_b_gravity_eqs} we 
solve gravity equations for the deformed conifold  and calculate the potential for the complex structure modulus $b$ in the 
large $b$ limit.  In Sec. \ref{sec:quarkmasses} we interpret $H_3$-form in terms of quark masses in 4D SQCD. We also discuss the  degeneration of the conifold at the runaway vacuum as a flow of 4D SQCD to  \ntwo supersymmetric QED (SQED) upon decoupling of two quark flavors. Sec. \ref{sec:conclusions} summarizes  our conclusions.

\section {Non-Abelian critical vortex string}
\label{sec:NAstring}
\setcounter{equation}{0}

\subsection{Four-dimensional \boldmath{${\mathcal N}=2\;$} 
 SQCD}
\label{sec:SQCD}

As was already mentioned, non-Abelian vortex-strings were first found in 4D
\ntwo SQCD with the gauge group U$(N)$ and $N_f \ge N$ quark flavors 
supplemented by the Fayet-Iliopoulos (FI)  term  \cite{FI} with parameter $\xi$
\cite{HT1,ABEKY,SYmon,HT2}, see for example, \cite{SYrev} for a detailed review of this theory.
Here, we just mention that at weak coupling $g^2\ll 1$, this theory is in the Higgs phase in which the scalar
components of the quark multiplets (squarks) develop vacuum expectation values (VEVs). These VEVs break 
the U$(N)$ gauge group
Higgsing  all gauge bosons. The Higgsed gauge bosons combine with the screened quarks to form long \ntwo multiplets with mass $m_G \sim g\sqrt{\xi}$.

 The global flavor SU$(N_f)$ is broken down to the so called color-flavor
locked group. The resulting global symmetry is
\beq
 {\rm SU}(N)_{C+F}\times {\rm SU}(N_f-N)\times {\rm U}(1)_B,
\label{c+f}
\eeq
see \cite{SYrev} for more details. 

The unbroken global U(1)$_B$ factor above is identified with a baryonic symmetry. Note that 
what is usually identified as the baryonic U(1) charge is a part of  our 4D theory  gauge group.
 ``Our" U(1)$_B$
is  an unbroken by squark VEVs combination of two U(1) symmetries;  the first is a subgroup of the flavor 
SU$(N_f)$, and the second is the global U(1) subgroup of U$(N)$ gauge symmetry.

As was already noted, we consider \ntwo SQCD  in the Higgs phase:  $N$ squarks  condense. Therefore,  non-Abelian 
vortex strings confine monopoles. In the \ntwo 4D theory these strings are 1/2 BPS-saturated; hence,  their
tension  is determined  exactly by the FI parameter,
\beq
T=2\pi \xi\,.
\label{ten}
\eeq
However, as we already mentioned, non-Abelian strings cannot be broken, therefore 
 monopoles cannot be attached to the string end points. In fact, in the U$(N)$ theories confined  
 monopoles 
are  junctions of two distinct elementary non-Abelian strings \cite{T,SYmon,HT2} (see \cite{SYrev} 
for a review). As a result,
in  four-dimensional \ntwo SQCD we have 
monopole-anti-monopole mesons in which the monopole and anti-monopole are connected by two confining strings.
 In addition, in the U$(N)$  gauge theory we can have baryons  appearing as  a closed 
``necklace'' configurations of $N\times$(integer) monopoles \cite{SYrev}. For the U(2) gauge group the 
massless BPS baryon $b$ found from  string theory in \cite{KSYconifold} consists of four monopoles \cite{ISY_b_baryon}.

Below we focus on the particular case $N=2$ and $N_f=4$ because, as was mentioned in the Introduction, in this case 4D \ntwo SQCD supports non-Abelian vortex strings which behave as critical superstrings \cite{SYcstring}.  Also, for $N_f=2N$ the gauge coupling $g^2$ of the 4D SQCD does not run; the $\beta$ function 
vanishes. However, the conformal invariance of the 4D theory is explicitly broken by the FI parameter $\xi$, which 
defines VEV's of quarks. The FI parameter is not renormalized.

Both stringy monopole-antimonopole mesons and monopole baryons with spins $J\sim 1$ have masses determined 
by the string tension,  $\sim \sqrt{\xi}$ and are heavier at weak coupling $g^2\ll 1$ than perturbative states with masses
$m_G\sim g\sqrt{\xi}$. 
Thus, they can decay into perturbative states \footnote{Their quantum numbers with respect to the global group 
\eqref{c+f} allow these decays, see \cite{SYrev}.} and in fact at weak coupling we do not 
expect them to appear as stable  states.

Only in the   strong coupling domain $g^2\sim 1$  we expect that (at least some of) stringy mesons and baryons become stable.
These expectations were confirmed in \cite{KSYconifold,SYlittles} where low lying string states in the string theory for the critical non-Abelian vortex were found at the self-dual point at strong coupling.

Below in this paper we introduce  quark masses $m_A$, $A=1,...4$ assuming that two first squark flavors with masses $m_1$ and $m_2$ develop VEVs.

\subsection{World-sheet sigma model}
\label{sec:wcp}

The presence of the color-flavor locked group SU$(N)_{C+F}$ is the reason for the formation of 
non-Abelian vortex strings \cite{HT1,ABEKY,SYmon,HT2}.
The most important feature of these vortices is the presence of the  orientational  zero modes.
As was already mentioned, in \ntwo SQCD these strings are 1/2 BPS saturated and preserve \ntwot supersymmetry on the world sheet. 

Let us briefly review the model emerging on the world sheet
of the non-Abelian  string \cite{SYrev}.

The translational moduli fields  are described by the Nambu-Goto action and  decouple from all other moduli. Below we focus on
 internal moduli.

If $N_f=N$  the dynamics of the orientational zero modes of the non-Abelian vortex, which become 
orientational moduli fields 
 on the world sheet, are described by 2D
\ntwot supersymmetric ${\mathbb{CP}}(N-1)$ model.

If one adds additional quark flavors, non-Abelian vortices become semilocal --
they acquire size moduli \cite{AchVas}.  
In particular, for the non-Abelian semilocal vortex in U(2) \ntwo SQCD with four flavors,  in 
addition to  the complex orientational moduli  $n^P$ (here $P=1,2$), we must add two complex size moduli   
$\rho^K$ (where $K=3,4$), see \cite{HT2,HT1,AchVas,SYsem,Jsem,SVY}. 

The effective theory on the string world sheet is a two-dimensional \ntwot supersymmetric \wcpt model, see review 
\cite{SYrev} for details. This model 
can be  defined  as a low energy limit of the  U(1) gauge theory \cite{W93}. The fields $n^{P}$ and $\rho^K$ have
charges  $+1$ and $-1$ with respect to the  U(1) gauge field. The  target space of the \wcpt model
is defined by the $D$-term condition
\beq
|n^{P}|^2-|\rho^K|^2 = {\rm Re}\,\beta, \qquad P=1,2, \qquad K=3,4.
\label{D-term}
\eeq
The number of real bosonic degrees of freedom in the model \wcpt is $8-1-1=6$.  Here 8 is the number of real degrees of 
freedom of $n^P$ and $\rho^K$ fields and we subtracted one real constraint imposed by the the $D$ term condition in 
\eqref{D-term}  and one  gauge phase eaten by the Higgs mechanism. As we already mentioned, these six internal degrees of freedom in the massless limit can be combined with four translational moduli to form a 10D space needed for a superstring to be critical. 

The global symmetry of the world sheet  \wcpt model is
\begin{equation}
	 {\rm SU}(2)\times {\rm SU}(2)\times {\rm U}(1)_B \,,
\label{globgroup}
\end{equation}
i.e. exactly the same as the unbroken global group in the 4D theory at $N=2$ and $N_f=4$. 
The fields $n$ and $\rho$ 
transform in the following representations:
\begin{equation}
	n:\quad \left(\textbf{2},\,\textbf{1},\, \frac12\right), \qquad \rho:\quad \left(\textbf{1},\,\textbf{2},\, \frac12\right)\,.
\label{repsnrho}
\end{equation}
Here  the global ``baryonic''  U(1)$_B$ group   rotates $n$ and 
$\rho$ fields with the same phase,  see \cite{KSYconifold} for details.

Twisted masses of $n^P$ and $\rho^K$ fields coincide with quark masses of 4D SQCD and are given respectively by $m_P$ and 
$m_K$, $P=1,2$ and $K=3,4$, see \cite{SYrev}.
Non-zero twisted masses $m_A$ break each of the SU(2) factors in \eqref{globgroup} down to U(1).

The 2D coupling constant ${\rm Re}\,\beta$ can be naturally complexified to the complex coupling constant $\beta$ if we
include the $\theta$ term in the action \cite{W93}.
At  the quantum level, the coupling $\beta$ does not run in this theory. Thus, 
the \wcpt  model is superconformal at zero masses $m_A = 0$. Therefore, its target space is Ricci flat and  (being K\"ahler due to \ntwot supersymmetry) represents  a non-compact Calabi-Yau manifold,  namely the conifold $Y_6$, see \cite{NVafa} for a review.

The \wcpt model  with $m_A=0$ was used in \cite{SYcstring,KSYconifold} to define 
the critical string theory for the non-Abelian vortex at hand.

Typically solitonic strings are ''thick'' and the effective world sheet theory  has 
a series of unknown high derivative corrections in powers of  $\pt/m_G$.
The string transverse size is given  by $1/m_G$, where $m_G \sim g\sqrt{\xi}$ is 
a  mass scale  of the gauge bosons and quarks  forming the string. The string cannot be thin in a  weakly
coupled 4D SQCD because at weak coupling $m_G\sim g\sqrt{T}$ and $m_G^2$ is always small in the units of the string tension $T$,
see \eqref{ten}.

A conjecture was put forward in  \cite{SYcstring}  that  at strong coupling 
in the vicinity of a critical value  $g_c^2\sim 1$ the non-Abelian string in the theory at hand  becomes thin,
and higher-derivative corrections in the world sheet theory  are absent. This is possible because the low energy 
\wcpt  model  already describes a critical string and higher-derivative corrections are not required to improve its 
ultra-violet behavior, see \cite{PolchStrom} for the discussion of this problem.
 The above  conjecture implies that $m_G(g^2) \to \infty$ at  $ g^2\to g_c^2$. As expected 
the thin string produces linear Regge trajectories even for small spins \cite{SYlittmult}.
 
It was also conjectured in \cite{KSYconifold} that $g_c$ corresponds to the value of the 2D coupling constant $\beta=0$.
The motivation for this conjecture is that  this value is a self-dual point for the \wcpt model. Also $\beta=0$ is a natural choice because at this point we have a regime change in the \wcpt model. The resolved conifold defined by the $D$ term condition \eqref{D-term} develops a conical  singularity at this point.  The point $\beta=0$ corresponds to 
$\tau_{SW} =1$ in the 4D SQCD, where $\tau_{SW}$ is the complexified inverse coupling, $\tau_{SW}= i\frac{8\pi}{g^2} 
+ \frac{\theta_{4D}}{\pi}$, where $\theta_{4D}$ is the 4D $\theta$ angle \cite{ISY_b_baryon}.

As we already mentioned in the Introduction a solitonic string-gauge duality proposed in \cite{SYcstring,KSYconifold} for 4D SQCD imply that
 at weak coupling this 
theory is in the Higgs phase and can be 
described in terms of quarks and Higgsed gauge bosons, while at strong coupling hadrons of this theory 
can be understood as closed string states in the string theory on $\mathbb{R}^4\times Y_6$.

Nonzero twisted masses $m_A\neq 0$ define  a mass deformation of the  superconformal CY theory
on the conifold. Generically  quark masses break the world sheet conformal invariance. The  \wcpt model with nonzero $m_A$   can no longer be used to define a string theory for the non-Abelian vortex in the massive 4D SQCD.

\subsection {Massless 4D baryon}
\label{conifold}

In this section we briefly review the only 4D massless state found in the string theory of the critical non-Abelian vortex
in the massless limit \cite{KSYconifold}. It is associated 
with the deformation of the conifold complex structure. 
 As was already mentioned, all other massless string modes  have non-normalizable wave functions over the conifold. In particular, 4D graviton associated with a constant wave
function over the conifold $Y_6$ is
absent \cite{KSYconifold}. This result matches our expectations since we started with
\ntwo SQCD in the flat four-dimensional space without gravity.

We can construct the U(1) gauge-invariant ``mesonic'' variables
\beq
w^{PK}= n^P \rho^K.
\label{w}
\eeq
These variables are subject to the constraint
\beq
{\rm det}\, w^{PK} =0. 
\label{coni}
\eeq

Equation (\ref{coni}) defines the conifold $Y_6$.  
It has the K\"ahler Ricci-flat metric and represents a non-compact
 Calabi-Yau manifold \cite{Candel,NVafa,W93}. It is a cone which can be parametrized 
by the non-compact radial coordinate 
\beq
\widetilde{r}^{\, 2} = {\rm Tr}\, \bar{w}w\,
\label{tilder}
\eeq
and five angles, see \cite{Candel}. Its section at fixed $\widetilde{r}$ is $S_2\times S_3$.

At $\beta =0$ the conifold develops a conical singularity, so both spheres $S_2$ and $S_3$  
can shrink to zero.
The conifold singularity can be smoothed out
in two distinct ways: by deforming the K\"ahler form or by  deforming the 
complex structure. The first option is called the resolved conifold and amounts to keeping
a non-zero value of $\beta$ in (\ref{D-term}). This resolution preserves 
the K\"ahler structure and Ricci-flatness of the metric. 
If we put $\rho^K=0$ in \eqref{D-term} we get the $\mathbb{CP}(1)$ model with the sphere $S_2$ as a target space
(with the radius $\sqrt{\beta}$).  
The resolved conifold has no normalizable zero modes. 
In particular, 
the modulus $\beta$  which becomes a scalar field in four dimensions
 has non-normalizable wave function over the 
$Y_6$ and therefore is not dynamical \cite{KSYconifold}.  

If $\beta=0$ another option exists, namely a deformation 
of the complex structure \cite{NVafa}. 
It   preserves the
K\"ahler  structure and Ricci-flatness  of the conifold and is 
usually referred to as the deformed conifold. 
It  is defined by deformation of Eq.~(\ref{coni}), namely,   
\beq
 {\rm det}\, w^{PK} = b\,,
\label{deformedconi}
\eeq
where $b$ is a complex parameter.
Now  the sphere $S_3$ can not shrink to zero, its minimal size is determined by $b$. 

The modulus $b$ becomes a 4D complex scalar field. The  effective action for  this field was calculated in \cite{KSYconifold}
using the explicit metric on the deformed conifold  \cite{Candel,Ohta,KlebStrass},
\beq
S_{{\rm kin}}(b) = T\int d^4x |\pt_{\mu} b|^2 \,
\log{\frac{\widetilde{R}_{\rm IR}^2}{|b|}}\,,
\label{Sb}
\eeq
where $\widetilde{R}_{\rm IR}$ is the  maximal value of the radial coordinate $\widetilde{r}$  introduced as an infrared regularization of the 
logarithmically divergent $b$-field  norm. Here the logarithmic integral at small $\widetilde{r}$ is cut off by the minimal size of $S_3$, which is equal to $|b|$.

To avoid confusion we  note that in AdS/CFT correspondence the radial coordinate of  internal dimensions has an interpretation of energy. The large values of this coordinate correspond to the ultraviolet region. In our approach it is vise-verse. The radial coordinate $\widetilde{r}$ measures absolute values of products $n^P\rho^K$ and since $\rho$'s are vortex string size moduli \cite{AchVas} $\widetilde{r}$ has a 4D interpretation as a distance from the string axis. In particular, large 
$\widetilde{r}$ corresponds to the infrared region.

We see that the norm of
the  modulus $b$ turns out to be  logarithmically divergent in the infrared.
The modes with the logarithmically divergent norm are at the borderline between normalizable 
and non-normalizable modes. Usually
such states are considered as ``localized'' in the 4D. We follow this rule. 
This scalar mode is localized near the conifold singularity  in the same sense as the orientational 
and size zero modes are localized on the vortex string solution, see \cite{SVY}.
   
 The field $b$  being massless can develop a VEV. Thus, 
we have a new Higgs branch in 4D \ntwo SQCD which is developed only for the critical value of 
the 4D coupling constant $\tau_{SW}=1$ associated with $\beta=0$.

 In \cite{KSYconifold} the massless state $b$ was interpreted as a baryon of 4D \ntwo QCD.
Let us explain this.
 From Eq.~(\ref{deformedconi}) we see that the complex 
parameter $b$ (which is promoted to a 4D scalar field) is a singlet with respect to both SU(2) factors in
 (\ref{globgroup}), i.e. 
the global world-sheet group.\footnote{Which is isomorphic to the 4D
global group \eqref{c+f} for $N=2$, $N_f=4$.} What about its baryonic charge? From \eqref{repsnrho} and \eqref{deformedconi}
we see that the $b$ state transforms as 
\beq
({\bf 1},\,{\bf 1},\,2).
\label{brep}
\eeq
 In particular it has the baryon charge $Q_B(b)=2$.

 In type IIA superstring compactifications the complex scalar 
associated with deformations of the complex structure of the Calabi-Yau
space enters as a 4D \ntwo BPS hypermultiplet, see \cite{Louis} for a review. 

On the field theory side we know that if we switch on generic quark masses in 4D SQCD the $b$-baryon becomes massive. 
Since it is a BPS state its mass is dictated by its baryonic charge \cite{ISY_b_baryon},
\beq
m_{b} = |m_1+m_2-m_3-m_4|.
\label{m_b}
\eeq

To conclude this section let us present the explicit metric of the singular conifold (with both $\beta$ and $b$ equal to zero), which will be used in the next section.
It has the  form  \cite{Candel}
\beq
ds^2_{6}=dr^2 + \frac{r^2}{6}(e_{\theta_1}^2+ e_{\varphi_1}^2 +e_{\theta_2}^2+ e_{\varphi_2}^2) +\frac{r^2}{9}e_{\psi}^2 ,
\label{conmet}
\eeq
where
\beqn
&& e_{\theta_1}= d\theta_1 , \qquad  e_{\varphi_1}= \sin{\theta_1}\, d\varphi_1\,,
\nonumber\\
&& e_{\theta_2}= d\theta_2 , \qquad  e_{\varphi_2}= \sin{\theta_2}\, d\varphi_2\,,
\nonumber\\
&& e_{\psi}= d\psi  + \cos{\theta_1}d\varphi_1+ \cos{\theta_2}d\varphi_2\,.
\label{angles}
\eeqn
Here $r$ is another  radial coordinate on the cone while the angles above are defined at $0\le \theta_{1,2}<\pi$,
$0\le \varphi_{1,2}<2\pi$, $0\le \psi<4\pi$.

\vspace{2mm}

The volume integral associated with this metric is
\beq
({\rm Vol})_{Y_6} = \frac{1}{108}\int r^5 \,dr \, d\psi\,  d\theta_1\,\sin{\theta_1} d\varphi_1 \,d\theta_2 \,
 \sin{\theta_2}d\varphi_2\,.
\label{Vol}
\eeq
The  radial coordinate, $\widetilde{r}$ defined in terms of matrix $w^{PK}$, see \eqref{tilder}  is related to
$r$ in (\ref{conmet}) via \cite{Candel}
\beq
r^2 = \frac32 \,\widetilde{r}^{4/3}\,.
\label{rtilder}
\eeq

\section {Gravity equations in the large $r$ limit}
\label{sec:gravity_eqs}
\setcounter{equation}{0}

Below we switch on NS 3-form flux $H_3$ and study its back reaction on the metric and the dilaton solving gravity equations of motion.
As we already mentioned in the Introduction $H_3$ flux produces a potential lifting the flat direction associated with 
the conifold complex structure modulus $b$. We confirm the result obtained in \cite{Y_NSflux} for this potential.

In this section we start with the large $r$ limit and show that the geometry is smooth and metric warp factors do not develop 
singularities at $r\to\infty$. Large $r$ limit means that $r\gg |b|^{1/3}$ (see \eqref{rtilder}) so for $H_3=0$  we can use the metric of the singular conifold \eqref{conmet}.

\subsection{The setup}

The bosonic part of the action of the type IIA supergravity in the Einstein frame is given by
\beqn
&& S_{10D} = \frac1{2\kappa^2}\,
\int d^{10} x \sqrt{-G}\left\{ R - \frac12 G^{MN}\,\pt_M\Phi\,\pt_N\Phi 
\right.
\nonumber\\
&& -
\left.
\frac{e^{-\Phi}}{12}\,H_{MNL}H^{MNL} 
\right\}, 
\label{10Daction}
\eeqn
where $G_{MN}$ and $\Phi$ are 10D metric and  dilaton, the string coupling $g_s=e^{\Phi}$. We also keep only NS 2-form $B_2$ with the field strength $H_3=dB_2$. We do not consider RR forms here, in particular,  the RR 3-form potential $C_3$. For compact CYs
the mass term for complex structure moduli can be generated via topological term $\int \frac12\, H_3\wedge C_3\wedge dC_3$
in the action \cite{Louis2}. However, it was shown in \cite{Y_NSflux} that for the noncompact case of the  conifold
this mechanism does not work due to the non-normalizability of the 4D part of $C_3$.

Einstein's equations of motion following from action \eqref{10Daction} have the form
\beq
R_{MN}= \frac12 \,\pt_M\Phi\,\pt_N\Phi + \frac{e^{-\Phi}}{4}\, H_{MAB}H_N^{AB} - \frac{e^{-\Phi}}{48}\,G_{MN}\, H_3^2,
\label{Einstein}
\eeq
while the equation for the dilaton reads
\beq
G^{MN}D_M\Phi D_N \Phi + \frac{e^{-\Phi}}{12}\, H_3^2 =0.
\label{dilatoneq}
\eeq
Finally the equation for the NS 3-form is
\beq
d(e^{-\Phi}\ast H_3) =0,
\label{H_3eqn}
\eeq
where $\ast $ denotes the  Hodge star.

We will see below that we need to introduce four warp factors to solve Einstein equations. Our ansatz for the metric is
\beq
ds^2_{10} = T\,h^{-1/2}_4 (r) \,\eta_{\mu\nu} dx^{\mu}dx^{\nu}+ \,g_{mn} dx^m dx^n,
\label{10met}
\eeq
where  $\mu,\nu =0,...,3$ are indices of the 4D space and $\eta_{\mu\nu}$ is the  flat Minkowski metric with signature
$(-1,1,1,1)$, while $m,n= 5,...10$ are indices  of the 6D internal space. Here internal coordinates $x^m$ defined to be dimensionless to match the dimension of scalar fields in the world sheet \wcpt model. We also introduced the string tension $T$ (see \eqref{ten}) in \eqref{10met} to fix dimensions.

The internal space  has a conifold metric deformed by  three warp factors
\beq
g_{mn} dx^m dx^n = h^{1/2}_6 (r) \left\{ a(r)\,dr^2 + \frac{r^2}{6}(e_{\theta_1}^2+ e_{\varphi_1}^2 +e_{\theta_2}^2+ 
e_{\varphi_2}^2) +\frac{r^2}{9} \omega(r)\,e_{\psi}^2\right\},
\label{warpedconi}
\eeq
see \eqref{conmet} and we assume that warp factors $h_4$, $h_6$, $a$ and $\omega$ depend only on the radial coordinate $r$.
If $H_3=0$ all warp factors are equal to unity and the 10D space has  the structure $\mathbb{R}^4\times Y_6$.

\subsection{NS 3-form at large $r$}

We will see below that  solution of  gravity equations of motion in the large $r$ limit can be expanded in powers of 
$\mu^2/r^4$, where $\mu$ parametrize the $H_3$ flux. To find its behavior  we can use a perturbation
 theory in powers of the above parameter. At the first step we solve equations of motion for $H_3$ form using undeformed conifold metric. This was done
in \cite{Y_NSflux}. 

Let us define two real 3-forms
on $Y_6$,
\beq
\alpha_3\equiv \frac{dr}{r}\wedge \left(e_{\theta_1}\wedge e_{\varphi_1} -   e_{\theta_2}\wedge e_{\varphi_2}\right)
\label{alpha}
\eeq
and 
\beq
\beta_3\equiv e_{\psi}\wedge \left(e_{\theta_1}\wedge e_{\varphi_1} -   e_{\theta_2}\wedge e_{\varphi_2}\right)
\label{beta}
\eeq
They are both closed \cite{KlebNekras,KlebTseytlin},
\beq 
d\alpha_3 =0, \qquad d\beta_3 =0,
\eeq
Moreover, using the conifold metric \eqref{conmet} to the leading order one can check that their 10D-duals are given by
\beqn
&& \ast\alpha_3 \approx -\frac{T^2}{3}\, dx^0 \wedge dx^1 \wedge dx^2 \wedge dx^3 \wedge\beta_3, 
\nonumber\\
&&\ast\beta_3 \approx  3\, T^2\, dx^0 \wedge dx^1 \wedge dx^2 \wedge dx^3 \wedge\alpha_3.
\label{alphabetadual}
\eeqn

The above relations ensure that both 10D-dual forms are also closed.
\beq
d\ast\alpha_3 =0, \qquad d\ast\beta_3 =0.
\label{eqnofmotion}
\eeq

Two solutions for $H_3$-form found in \cite{Y_NSflux} are
\beq
H_3 \approx \mu_1 \alpha_3 + \frac{\mu_2}{3} \,\beta_3,
\label{H_3sol}
\eeq
where $\mu_1$ and $\mu_2$ are two independent real parameters, while the factor $\frac13$ is introduced for convenience.
 This $H_3$-form satisfy both Bianchi identity and equations of motion \eqref{H_3eqn}, where the dilaton is considered as a constant to the leading order in $\mu^2/r^4$.

3-Forms \eqref{alpha} and \eqref{beta} form a basis similar to the simplectic basis of harmonic $\alpha$ and $\beta$ 3-forms for compact CYs, see for example review \cite{Louis}. In particular,
\beq
\int_{Y_6} \alpha_3 \wedge \alpha_3 = \int_{Y_6} \beta_3 \wedge \beta_3 = 0,
\label{aabb}
\eeq
while 
\beq
\int_{Y_6} \alpha_3 \wedge \beta_3 \sim -\int \frac{dr}{r} \sim - \log{\frac{R_{\rm IR}^3}{|b|}}. 
\label{ab}
\eeq
Here $R_{\rm IR}$ is the maximal value of the radial coordinate $r$ introduced to regularize the infrared logarithmic divergence,
while at small $r$ the integral is cut off by the minimal size of $S_3$ which is equal to $|b|$.  Note that this logarithm is similar to the one, which determines the metric
for the $b$-baryon in \eqref{Sb} \footnote{Note that $R_{\rm IR}^3\sim \widetilde{R}_{\rm IR}^2$, see \eqref{rtilder}.}.

\subsection{Warp factors at large $r$}
\label{sec_warp_r}

For Minkowski indices $\mu,\nu= 0,1,2,3$ Einstein's  equations \eqref{Einstein} read
\beq
R_{\mu\nu} = - \frac{\eta_{\mu\nu}}{48}\, \frac{e^{-\Phi}}{h_4^{1/2}}\, H_3^2,
\label{munu_initial}
\eeq
where Ricci components for the ansatz  \eqref{10met}, \eqref{warpedconi} can be calculated using results of \cite{deWit}
\beq
R_{\mu\nu}=\frac{\eta_{\mu\nu}}{4ah_4^{1/2}h_6^{1/2}}\left\{\frac1{h_4} \Delta h_4 +\frac{h_6'h_4'}{h_6 h_4} - 2\frac{(h_4')^2}{h_4^2} -\frac12\,\frac{a'h_4'}{a h_4} + \frac12\,\frac{\omega'h_4'}{\omega h_4}\right\}.
\label{munu_comp}
\eeq
Here prime denotes the derivative with respect to $r$ and $\Delta$ is the Laplacian calculated using   the conifold metric \eqref{conmet}. 

Using expression in \eqref{munu_comp} we can compare Einstein's equations for Minkowski indices \eqref{munu_initial} with the dilaton equation \eqref{dilatoneq}. Rewriting the latter one as 
\beq
\Delta \Phi +\left( \frac{h_6'}{h_6} -  \frac{h_4'}{h_4} -\frac12\,\frac{a'}{a} + \frac12\,\frac{\psi'}{\omega}\right)\Phi'
 = -  \frac{e^{-\Phi}}{12}\, ah_6^{1/2}\,H_3^2
\eeq
it is easy to see that it is identical to the equation \eqref{munu_initial} upon substitution
\beq
\Phi = \Phi_0 + \ln{h_4},
\label{dilaton_solution}
\eeq
where $\Phi_0$ is a constant value of the dilaton present at $H_3=0$.

Let us now continue studying the equation \eqref{munu_initial}. 
At the first non-trivial order in the parameter $\mu^2/r^4$ all non-linearities in the expression in \eqref{munu_comp} can be neglected and it reduces simply to 
\beq
R_{\mu\nu}\approx \frac{\eta_{\mu\nu}}{4}\, \Delta h_4.
\eeq
This gives for the Minkowski part of Einstein's equations
\beq
\Delta h_4 \approx -\frac{e^{-\Phi_0}}{12}\, H_3^2,
\label{munu_eqs}
\eeq
where $H_3^2$ can be  calculated using the conifold metric and we used only the constant part of the dilaton $\Phi_0$ at this order.  We have 
\beq
e^{-\Phi_0}\, H_3^2 =3! \,72 \,\frac{\mu_1^2+\mu_2^2}{g_s}\,\frac1{r^6} = 2^4\,3^3\,\frac{\mu_1^2+\mu_2^2}{g_s}\,\frac1{r^6}
\label{H_3^2}
\eeq
where $g_s=e^{\Phi_0}$, while $72/r^4$ say, for the first solution for $H_3$ (proportional to $\mu_1$ in \eqref{H_3sol}) comes from $g^{\theta_1\theta_1}g^{\varphi_1\varphi_1}$ and 
$g^{\theta_2\theta_2}g^{\varphi_2\varphi_2}$.

Then equation \eqref{munu_eqs} gives
\beq
h_4=1+ \frac{9}{g_s}\,\frac{\mu_1^2+\mu_2^2}{r^4}\,\log{\frac{r}{|b|^{1/3}}} + O(\mu^4/r^8).
\label{h_4sol}
\eeq
up to a non-logarithmic term proportional to $\mu^2/r^4$ which we set to zero.

Consider now Einstein's equations with internal indices.
Let  index $\alpha$ ($\beta$) denote differentials $e_{\theta_1},e_{\varphi_1},e_{\theta_2},e_{\varphi_2}$.
Then we can calculate Christoffel symbols with $r$ indices, namely
\beqn
&& \Gamma^{r}_{\alpha\beta}= -\frac{g^{(c)}_{\alpha\beta}}{a}\left(\frac1r+\frac14\,\frac{h_6'}{h_6}\right) , \qquad  
\Gamma^{r}_{\psi\psi}= -\frac{g^{(c)}_{\psi\psi}}{a}\left(\frac1r+\frac14\,\frac{h_6'}{h_6} +\frac12\omega'\right)\,,
\nonumber\\
&&\Gamma^{\beta}_{r\alpha}=\Gamma^{\beta}_{\alpha r}=\delta^{\beta}_{\alpha}\left(\frac1r+\frac14\,\frac{h_6'}{h_6}\right),
\qquad \Gamma^{\psi}_{r\psi}=\Gamma^{\psi}_{\psi r}=\frac1r+\frac14\,\frac{h_6'}{h_6}+\frac12\omega'\,,
\nonumber\\
&&\Gamma^{r}_{rr}=\frac12\,\frac{a'}{a}+\frac14\,\frac{h_6'}{h_6}, \qquad \Gamma^{n}_{rr}= \Gamma^{r}_{rn}= \Gamma^{r}_{nr}=0,
\qquad n\neq r,
\label{Gammas_r}
\eeqn
where   $g^{(c)}_{\alpha\beta}$ and $g^{(c)}_{\psi\psi}$ denote the conifold metric \eqref{conmet}.

Using these formulas we find nonzero Ricci  components at the leading order in $\mu^2/r^4$. We have 
\beqn
&& R_{\alpha\beta}\approx g^{(c)}_{\alpha\beta}\left\{ \frac4{r^2}\,(a-1) -\frac14 \Delta h_6 -\frac1r\, h_6' +\frac1r\, h_4' +\frac1{2r}\,a'
-\frac1{2r}\,\omega'\right\},
\nonumber\\
&& R_{\psi\psi}\approx g^{(c)}_{\psi\psi}\left\{ \frac4{r^2}\,(a-1) -\frac14 \Delta h_6 -\frac1r\, h_6' +\frac1r\, h_4' +\frac1{2r}\,a'
-\frac2{r}\,\omega' -\frac12\,\omega''\right\}, 
\nonumber\\
&& R_{rr}\approx   -\frac14 \Delta h_6 - h_6'' + h_4'' +\frac5{2r}\,a'
-\frac1{r}\,\omega' -\frac12\,\omega'' \, .
\label{Ricci_internal}
\eeqn
 Here we used that Ricci tensor is zero if all warp factors are equal to unity and the dependence on $h_4$ can be found using formulas in \cite{deWit}. 

Now calculating r.h.s.'s for Einstein's equations \eqref{Einstein} we get for the first solution proportional to $\mu_1$
in \eqref{H_3sol}
\beqn
&& R_{\alpha\beta}= \frac1{ 48}\,g^{(c)}_{\alpha\beta} e^{-\Phi_0}\, (H_3^{(1)})^2,
\nonumber\\
&& R_{\psi\psi}= -\frac{g^{(c)}_{\psi\psi}}{48}\, e^{-\Phi_0}\, (H_3^{(1)})^2
\nonumber\\
&&R_{rr}= \frac1{16}\,e^{-\Phi_0}\, (H_3^{(1)})^2,
\label{Einstein_eq_r1}
\eeqn
where $(H_3^{(1)})^2$ is given by \eqref{H_3^2} with $\mu_2=0$.

For the second solution in \eqref{H_3sol} (proportional to $\mu_2$ ) we have
\beqn
&& R_{\alpha\beta}= \frac1{ 48}\,g^{(c)}_{\alpha\beta} e^{-\Phi_0}\, (H_3^{(2)})^2,
\nonumber\\
&& R_{\psi\psi}= \frac{g^{(c)}_{\psi\psi}}{16}\,
 e^{-\Phi_0}\, (H_3^{(2)})^2
\nonumber\\
&&R_{rr}= -\frac1{48}\,e^{-\Phi_0}\, (H_3^{(2)})^2,
\label{Einstein_eq_r2}
\eeqn 
where $(H_3^{(2)})^2$ is given by \eqref{H_3^2} with $\mu_1=0$.

Above equations together with expressions \eqref{Ricci_internal} and solution for 
$h_4$ \eqref{h_4sol} determine three warp factors $h_6$, $a$ and $\omega$ at the leading order. For the first solution for 
$H_3$ we have
\beqn
&& h_6^{(1)}= 1+ \frac{9}{g_s}\,\frac{\mu_1^2}{r^4}\,\log{\frac{r}{|b|^{1/3}}} -\frac9{5}\,\frac1{g_s}\,\frac{\mu_1^2}{r^4}
+\cdots,
\nonumber\\
&& a^{(1)}=1-\frac9{10}\,\frac1{g_s}\,\frac{\mu_1^2}{r^4}+\cdots,
\nonumber\\
&& \omega^{(1)} = 1+\frac9{2}\,\frac1{g_s}\,\frac{\mu_1^2}{r^4} +\cdots,
\label{warp_sol_r1}
\eeqn
where dots stand for sub-leading terms of order of $\mu^4/r^8$. Warp factors for the second solution for $H_3$ have the form
\beqn
&& h_6^{(2)}= 1+ \frac{9}{g_s}\,\frac{\mu_2^2}{r^4}\,\log{\frac{r}{|b|^{1/3}}} +\frac9{5}\,\frac1{g_s}\,\frac{\mu_2^2}{r^4}
+\cdots,
\nonumber\\
&& a^{(2)}=1+\frac9{10}\,\frac1{g_s}\,\frac{\mu_2^2}{r^4}+\cdots,
\nonumber\\
&& \omega^{(2)} = 1-\frac9{2}\,\frac1{g_s}\,\frac{\mu_2^2}{r^4} +\cdots.
\label{warp_sol_r2}
\eeqn

Finally solutions \eqref{dilaton_solution} and \eqref{h_4sol}
give for the dilaton
\beq
e^{(\Phi- \Phi_0)} = 1+ \frac{9}{g_s}\,\frac{\mu_1^2+\mu_2^2}{r^4}\,\log{\frac{r}{|b|^{1/3}}} +\cdots
\label{dilaton_sol_r}
\eeq

We see that warp factors and dilaton have smooth behavior at large $r$ and can be found order by order in the parameter 
$\mu^2/r^4$ using perturbation theory in gravity equations. The region of validity of the above solutions is 
\beq
r\gg |b|^{1/3}\gg \mu^{1/2}.
\label{r_region}
\eeq

To conclude this section we would like to comment on a subtlety in solving equations \eqref{Einstein_eq_r1} and 
\eqref{Einstein_eq_r2}. In fact 
these equations do not determine coefficients in non-logarithmic terms proportional to $1/r^4$ for $h_6$ and $a$ separately.
Denoting these coefficients $\chi$ and $A$ respectively we find that the first and the third equations in \eqref{Einstein_eq_r1}
 and \eqref{Einstein_eq_r2} give the same conditions for them, namely
\beq
\chi^{(1)} +\frac{A^{(1)}}{2}= -\frac94\,\frac{\mu_1^2}{g_s}, \qquad \chi^{(2)} +\frac{A^{(2)}}{2}= \frac94\,\frac{\mu_2^2}{g_s}
\label{A_delta_eq}
\eeq
for \eqref{Einstein_eq_r1} and \eqref{Einstein_eq_r2} respectively.
The resolution of this puzzle is related to the possibility of redefinition of the conifold radial coordinate $r$. Let us put
$H_3=0$ so the metric is reduced to the conifold one in \eqref{conmet}. However, we can redefine $r$ at the relevant order,
\beq
r=f(r')= r'\left(1+\frac{\alpha}{r'^4}\right),
\eeq 
where $\alpha$ is a constant.
This gives 
\beq
r^2\approx r'^2\left(1+\frac{2\alpha}{r'^4}\right),\qquad dr^2\approx dr'^2\left(1-\frac{6\alpha}{r'^4}\right)
\eeq
which in terms of the new coordinate $r'$ imply nontrivial  warp factors
\beq
h_6 =1+\frac{4\alpha}{r'^4}, \qquad a= 1-\frac{8\alpha}{r'^4}
\eeq
or nonzero coefficients
\beq
\chi=4\alpha, \qquad A=-8\alpha.
\eeq

Now we see that the combination which enters Eqs. \eqref{A_delta_eq} is zero on this solution,
\beq
\chi + \frac{A}{2}=0.
\label{r_redef}
\eeq
Thus, nontrivial solutions of the above equation are related to the possibility of $r$ redefinition. 

To fix the definition of $r$ we require that the orthogonal combination to the one which enters \eqref{r_redef} should be zero, namely
\beq
\frac{\chi}{2} -A =0.
\label{fix_r}
\eeq
This condition together with equation \eqref{A_delta_eq} gives coefficients
\beqn
&&\chi^{(1)}= -\frac95\,\frac{\mu_1^2}{g_s}, \qquad A^{(1)}= -\frac9{10}\,\frac{\mu_1^2}{g_s},
\nonumber\\
&&\chi^{(2)}= \frac95\,\frac{\mu_2^2}{g_s}, \qquad A^{(2)}= \frac9{10}\,\frac{\mu_2^2}{g_s}
\eeqn
for two solutions for $H_3$ respectively, which we presented in \eqref{warp_sol_r1} and \eqref{warp_sol_r2}.

\subsection{The scalar potential}
\label{sec:pot}

To find the scalar potential induced by 3-form flux $H_3$ we substitute the solution of the gravity equations found 
above into the 10D action \eqref{10Daction}. The trace of the Einstein's equations \eqref{Einstein} reads
\beq
 R - \frac12 G^{MN}\,\pt_M\Phi\,\pt_N\Phi -\frac{e^{-\Phi}}{12}\,H_3^2 =0.
\label{R_10}
\eeq
Substituting this into Eq. \eqref{10Daction} we get the action calculated on the solution,
\beq
S_{10D}= \frac1{2\kappa^2}\,\int d^{10} x\sqrt{-G}\left\{-\frac{e^{-\Phi}}{24}\,H_3^2 \right\},
\eeq
where $2\kappa^2= (2\pi)^3g_s^2$ in our conventions.

This leads to the potential for $b$-baryon (complex structure modulus $b$ of the conifold) in 4D SQCD,
\beq
V(b)=\frac{T^2}{(2\pi)^3g_s^2}\,\int d^6 x \sqrt{g_6}\;\frac{e^{-\Phi}}{24}\,H_3^2 ,
\label{pot_gen}
\eeq
where the string tension $T$ appears due to our normalization of the Minkowski part of the metric, see \eqref{10met}. Here the integral is taken over the internal 6D space and $g_6$ is the determinant of the 6D metric.
To the leading order we can neglect warp factors and calculate the above integral using the conifold metric \eqref{conmet}.
Using Eqs. \eqref{Vol} and \eqref{H_3^2} we get
\beq
V(b) = \frac43\,\frac{T^2}{g_s^3}(\mu_1^2+\mu_2^2) \int \frac{dr}{r} = \frac49\,\frac{T^2}{g_s^3}(\mu_1^2+\mu_2^2)\,
\log{\frac{R^3_{IR}}{|b|}},
\label{pot_r}
\eeq
where  $R_{IR}$ is the infrared cutoff for the radial coordinate $r$,
while modulus $b$ plays a role of the ultraviolet cutoff at small $r$, cf. \eqref{ab}. This potential was calculated in 
\cite{Y_NSflux}.
Note, that the same  infrared logarithm
determines the metric  \eqref{Sb} for the $b$-baryon. If we take into account warp factors in the integrand in \eqref{pot_gen} this would give finite corrections to the potential of order of 
\beq
T^2\,\frac{\mu^4}{|b|^{4/3}},
\label{pot_cor_r}
\eeq
which are negligible as compared to the logarithmic term.

We see that the Higgs branch for $b$ is lifted by $H_3$ flux deformation and we have a runaway vacuum with VEV 
\beq
\bra|b|\ket \to R_{\rm IR}^3 \to \infty. 
\label{bVEV}
\eeq
However, our solution of gravity equations  is found in this section using the metric of the singular conifold and therefore is valid  at $r\gg |b|^{1/3}$.
Thus, the potential \eqref{pot_r} cannot be trusted    at 
$|b|\sim R_{\rm IR}^3$ where the logarithm becomes small. In the next section we consider the region of  $r\sim |b|^{1/3}$  and confirm our conclusion in \eqref{bVEV} that the VEV of the baryon $b$ tends to infinity.

\section {Gravity equations for the deformed conifold}
\label{sec:large_b_gravity_eqs}
\setcounter{equation}{0}

The result for the potential \eqref{pot_r} suggests that we have a runaway vacuum  and VEV of $b$ becomes infinitely large. To confirm this in this section we study gravity equations with nonzero $H_3$-flux on the deformed conifold assuming that the radial coordinate $r\sim |b|^{1/3}$. Anticipating the runaway behavior \eqref{bVEV} we still keep the second condition in 
\eqref{r_region},
\beq
\mu\ll |b|^{2/3}
\label{mu_condition}
\eeq

\subsection{Metric of the deformed conifold}

In this section we briefly review the metric of the deformed conifold.
It  has the form \cite{Candel,Ohta,KlebStrass}
\beqn
ds_6^2 &=& \frac12\,|b|^{2/3}\,K(\tau)\left\{ \frac{1}{3K^3(\tau)}\left(d \tau^2 + e_{\psi}^2\right)
+ \cosh^2{\frac{\tau}{2}}\,\left(g_3^2+ g_4^2\right) 
\right.
\nonumber \\
&+& \left.
 \sinh^2{\frac{\tau}{2}}\,\left(g_1^2+ g_2^2\right)\right\},
\label{defconmet}
\eeqn
where angle differentials are defined as 
\beqn
&&
g_1= -\frac1{\sqrt{2}}\,(e_{\phi_1} + e_3), \qquad g_2= \frac1{\sqrt{2}}\,(e_{\theta_1} - e_4),
\nonumber\\
&&
g_3= -\frac1{\sqrt{2}}\,(e_{\phi_1} - e_3), \qquad g_4= \frac1{\sqrt{2}}\,(e_{\theta_1} + e_4),
\label{g_angles}
\eeqn
while 
\beq
e_3= \cos{\psi}\sin{\theta_2}\, d\varphi_2 - \sin{\psi}\,d\theta_2, \qquad 
e_4= \sin{\psi}\sin{\theta_2}\, d\varphi_2 + \cos{\psi}\,d\theta_2,
\eeq
see also \eqref{angles}.

Here
\beq
K(\tau)=\frac{(\sinh{2 \tau}-2 \tau)^{1/3}}{2^{1/3}\sinh{\tau}}
\eeq 
and the new radial coordinate $\tau$ is defined as
\beq
\widetilde{r}^2=|b|\,\cosh{\tau} = \left(\frac23\right)^{\frac32}\,r^3.
\label{tau}
\eeq
In the limit of large $\tau$ the metric \eqref{defconmet} reduces to the metric \eqref{conmet} of the singular conifold.

Results of the previous section show that we have a runaway vacuum with $|b|\sim R_{IR}^3$ so we are interested in the metric
\eqref{defconmet} in the limit of small $\tau$, $\tau\ll 1$. In this limit the metric of the deformed conifold takes the form
\beqn
ds_6^2|_{\tau\to 0} &=& \frac12\,|b|^{2/3}\left(\frac23\right)^{\frac13}\,\left\{ \frac12\,d \tau^2 + \frac12\,e_{\psi}^2
+ g_3^2+ g_4^2 
\right.
\nonumber \\
&+& \left.
 \frac{\tau^2}{4}\,\left(g_1^2+ g_2^2\right)\right\}.
\label{conmettau0}
\eeqn
The last term here corresponds to the collapsing sphere $S_2$, while the sphere $S_3$ associated with three  angular terms 
in the first line  has a fixed radius in the limit $\tau\to 0$ \cite{Candel,KlebStrass}. The radial coordinate $r$ approaches its minimal value with
\beq
r^3|_{min} = \left(\frac32\right)^{\frac32}\, |b|
\label{r_min}
\eeq
at $\tau=0$.

The square root of the determinant of the metric
\beq
\sqrt{g_6} \sim |b|^2\,\cosh^2{\frac{\tau}{2}}\sinh^2{\frac{\tau}{2}} |_{\tau\to 0} \sim |b|^2\, \tau^2
\label{det}
\eeq
vanishes at $\tau=0$, which  shows the degeneration of the conifold metric.

\subsection{NS 3-form at small $\tau$}

We will see below that leading non-trivial contributions to warp factors are proportional to $\mu^2\tau^2/|b|^{4/3}$.
At the first step of the perturbation theory we can neglect them and look for solutions for $H_3$ flux using the metric of the deformed conifold summarized in the previous section and a constant dilaton, $\Phi\approx \Phi_0$.

One solution was found in \cite{Y_NSflux} using the ansatz  suggested in \cite{KlebStrass} for the type IIB flux compactification on the deformed conifold. The ansatz reads
\beq
H_3 = p'\, d\tau\wedge g_1\wedge g_2 + k'\, d\tau\wedge g_3\wedge g_4 - \frac12\,(p-k)\, e_{\psi}\wedge 
(g_1\wedge g_3 + g_2\wedge g_4),
\label{H_3tau1}
\eeq
where $p$ and $k$ are functions of the radial coordinate $\tau$. Here primes denote derivatives with respect to $\tau$.
The 3-form above is closed so the Bianchy identity is satisfied.

At large $\tau$ $p'\approx k'\to \mu_1/3$ and using the identity \cite{KlebStrass}
\beq
e_{\theta_1}\wedge e_{\varphi_1} - e_{\theta_2}\wedge e_{\varphi_2}= g_1\wedge g_2 + g_3\wedge g_4.
\label{e_g}
\eeq
it is easy to show that this solution tends to the first solution for $H_3$ (proportional to $\mu_1$) in \eqref{H_3sol}.

For small $\tau$ equation  of motion \eqref{H_3eqn} for $H_3$  was solved in \cite{Y_NSflux} at the leading order using the  metric of the deformed conifold and a constant dilaton. The result is   $k\approx \mu_1\tau$ and $p\approx -\mu_1\tau^5/80$ so the solution  takes the form
\beq
H_3^{(1)} \approx \mu_1 \gamma_3
\label{1st_sol_tau}
\eeq
 up to an overall constant, where we introduced a 3-form
\beq
\gamma_3= d\tau\wedge g_3\wedge g_4 -\frac{\tau^4}{16}\,d\tau\wedge g_1\wedge g_2 +\frac{\tau}{2}\, e_{\psi}\wedge 
( g_1\wedge g_3 +g_2\wedge g_4).
\label{gamma_3}
\eeq 

Now let us find another solution which at large $\tau$ tends to the second solution in \eqref{H_3sol} (proportional to $\mu_2$).
To do so we use the ansatz,
\beq
H_3= l(\tau)\, e_{\psi}\wedge g_1\wedge g_2 + n(\tau)\, e_{\psi}\wedge g_3\wedge g_4 + q(\tau)\, d\tau\wedge (g_1\wedge g_3 
+ g_2\wedge g_4),
\label{H_3tau2}
\eeq
where $l$, $n$ and $q$ are functions of $\tau$. Using identity \eqref{e_g} and \cite{KlebStrass}
\beq
d(g_1\wedge g_3 + g_2\wedge g_4) =  e_{\psi}\wedge (g_1\wedge g_2 - g_3\wedge g_4)
\label{id2}
\eeq
we calculate
\beq
dH_3=  l'\, d\tau\wedge e_{\psi}\wedge g_1\wedge g_2 + n'\,d\tau \wedge e_{\psi}\wedge g_3\wedge g_4 -
 q(\tau)\, d\tau\wedge e_{\psi}\wedge (g_1\wedge g_2 - g_3\wedge g_4).
\label{dH_3}
\eeq
Now Bianchy identity $dH_3=0$ leads to
\beq
l'-q=0, \qquad n'+q=0.
\label{l_n_eqs}
\eeq

A solution to these equations with $q=0$, $l=n=\mu_2/3$ corresponds  to the second solution in \eqref{H_3sol} at large $\tau$.
Let us find the extrapolation of this solution to small $\tau$. For nonzero $q$ we have $l'=-n'$ and setting the integration constant to zero we get $l=-n$. The ansatz for $H_3$ acquires the form
\beq
H_3= l\, \left(e_{\psi}\wedge g_1\wedge g_2 - e_{\psi}\wedge g_3\wedge g_4 \right) + l'\, d\tau\wedge (g_1\wedge g_3 
+ g_2\wedge g_4).
\label{H_3_l_only}
\eeq

Calculating 10D dual of \eqref{H_3_l_only} using metric in \eqref{conmettau0} we get 
\beqn
\ast H_3 &=& - T^2\, dx^0 \wedge dx^1 \wedge dx^2 \wedge dx^3 \wedge \left\{\frac{4l}{\tau^2}\, d\tau\wedge g_3\wedge g_4   
\right.
\nonumber\\
&-&
\left.
 \frac{l\tau^2}{4}\, d\tau\wedge g_1\wedge g_2
+l'\, e_{\psi}\wedge (g_1\wedge g_3 + g_2\wedge g_4) \right\}. 
\label{astH_3tau}
\eeqn

Then the equation of motion \eqref{H_3eqn} reads
\beqn
&&d\ast H_3 =  T^2\, dx^0 \wedge dx^1 \wedge dx^2 \wedge dx^3 \wedge 
\nonumber\\
&&
\left\{-\left(\frac{2l}{\tau^2}+\frac{l\tau^2}{8}-l''\right)\,d\tau\wedge e_{\psi}\wedge (g_1\wedge g_3 + g_2\wedge g_4) \right\} =0
\label{final_eq_H_3}
\eeqn
where we used the identity \cite{KlebStrass}
\beq
d(g_1\wedge g_2 - g_3\wedge g_4) = - e_{\psi}\wedge (g_1\wedge g_3 + g_2\wedge g_4).
\label{id1}
\eeq

Equation \eqref{final_eq_H_3} gives 
\beq
l''-\frac{2l}{\tau^2}=0,
\label{l_eq}
\eeq
where we neglect $\tau^2$-term at small $\tau$.

Eq. \eqref{l_eq} gives $l\approx \mu_2\tau^2/4$ up to a constant and we write down the second solution for $H_3$ in the form
\beq
H_3^{(2)}\approx \mu_2\delta_3,
\label{2nd_sol_tau}
\eeq
where
\beq
\delta_3= \frac{\tau^2}{4}\,e_{\psi}\wedge \left(g_1\wedge g_2 -  g_3\wedge g_4\right) +\frac{\tau}{2}\, 
d\tau\wedge ( g_1\wedge g_3 +g_2\wedge g_4).
\label{delta_3}
\eeq 

Both 3-forms $\gamma_3$ and $\delta_3$ are closed. Moreover, their 10D duals are given by (see \cite{Y_NSflux} 
and \eqref{astH_3tau})
\beqn
&& \ast\gamma_3 \approx T^2\, dx^0 \wedge dx^1 \wedge dx^2 \wedge dx^3 \wedge\delta_3, 
\nonumber\\
&&\ast\delta_3 \approx  - T^2\, dx^0 \wedge dx^1 \wedge dx^2 \wedge dx^3 \wedge\gamma_3.
\label{gamma_delta_dual}
\eeqn
 
The above relations ensure that both 10D-dual forms are also closed.
\beq
d\ast\gamma_3 =0, \qquad d\ast\delta_3 =0.
\label{closed_dual}
\eeq

Much in the same way as forms \eqref{alpha} and \eqref{beta} 3-forms \eqref{gamma_3} and \eqref{delta_3} satisfy relations
\beq
\int_{Y_6} \gamma_3 \wedge \gamma_3 = \int_{Y_6} \delta_3 \wedge \delta_3 = 0,
\label{ggdd}
\eeq
while 
\beq
\int_{Y_6} \gamma_3 \wedge \delta_3 \sim \int d\tau \tau^2 
\label{gd}
\eeq
at small $\tau$.

To conclude this section, we note that at $\tau=0$ the first solution  solution \eqref{1st_sol_tau} tends to a constant
\beq
H_3^{(1)}(\tau=0) = \mu_1\,d\tau\wedge g_3\wedge g_4,
\label{bc1}
\eeq
which we impose as  boundary conditions at $S_3$, which does not shrinks at $\tau=0$.  These boundary conditions ensure a non-zero solution for $H^{(1)}_3$.

Similarly for the second solution \eqref{2nd_sol_tau} we fix its derivative with respect to $\tau$ as boundary conditions 
at $S_3$ at $\tau=0$,
\beq
\frac{\pt}{\pt\tau}\,H_3^{(2)}(\tau=0) = \frac{\mu_2}{2}\,d\tau\wedge ( g_1\wedge g_3 +g_2\wedge g_4).
\label{bc2}
\eeq

\subsection{Warp factors at small $\tau$}

In this section we study the back reaction of the two solutions for $H_3$ flux found above on the metric and dilaton
to the leading order in $\mu^2\tau^2/|b|^{4/3}$. Our ansatz for the metric is given by \eqref{10met} where $h_4$ now is a function of $\tau$, while
\beqn
g_{mn} dx^m dx^n &=& \frac12\,|b|^{2/3}\left(\frac23\right)^{\frac13}\,\left\{ h_1^{1/2}(\tau)\,\left(\frac12\,a(\tau)
\,d \tau^2 + \frac12\,e_{\psi}^2 + g_3^2+ g_4^2\right) 
\right.
\nonumber \\
&+& \left.
 h_2^{1/2}(\tau)\,\frac{\tau^2}{4}\,\left(g_1^2+ g_2^2\right)\right\},
\label{warp_met_tau}
\eeqn
where the metric of the deformed conifold \eqref{conmettau0} is further deformed with another three warp factors $h_1$, $h_2$ and $a$,
which are assumed to be functions of $\tau$. Here we also assume the limit of small $\tau$, $\tau\ll 1$.

For Minkowski indices $\mu,\nu=0,1,2,3$ Einstein's equations \eqref{Einstein} has the form \eqref{munu_initial}, where
using results from \cite{deWit} we calculate
\beq
R_{\mu\nu}=\frac{\eta_{\mu\nu}\,g^{\tau\tau}_{c}}{4ah_4^{1/2}h_1^{1/2}}\left\{\frac1{h_4} \Delta h_4 +\frac12\frac{h_1'h_4'}{h_1 h_4}  +\frac12\frac{h_2'h_4'}{h_2 h_4}- 2\frac{(h_4')^2}{h_4^2} -\frac12\,\frac{a'h_4'}{a h_4} \right\},
\label{munu_comp_tau}
\eeq
where $\Delta$ is the Laplacian calculated using metric \eqref{conmettau0}. Here and below $g^{mn}_{c}$, $g_{mn}^{c}$ denote the deformed conifold metric \eqref{conmettau0}, for example
\beq
g^{\tau\tau}_{c} \approx \frac{2^{5/3}3^{1/3}}{|b|^{2/3}}.
\label{g^tautau}
\eeq

At the first order all non-linearities in \eqref{munu_comp_tau} can be neglected and Einstein's equations \eqref{munu_initial}
reduce to
\beq
\Delta h_4 \approx -\frac{e^{-\Phi_0}\,g_{\tau\tau}^{c}}{12}\, H_3^2,
\label{munu_eqs_tau}
\eeq
where $H_3^2$ can be  calculated using the deformed conifold metric and we used only the constant part of the dilaton $\Phi_0$ at this order.  We have 
\beq
e^{-\Phi_0}\, H_3^2 \approx 2^4\,3^3 \,\frac{\mu_1^2+\mu_2^2}{g_s}\,\frac1{|b|^{2}}\,\left[1 + O(\tau^2)\right] 
\label{H_3^2_tau}
\eeq
where,  say, for the first solution for $H_3$ in \eqref{1st_sol_tau} only first and the last terms in $\gamma_3$ contribute at the leading order in $\tau$.

Then equation \eqref{munu_eqs_tau} gives
\beq
h_4=1- \frac{3^{2/3}}{2^{2/3}}\,\frac{\mu_1^2+\mu_2^2}{g_s}\,\frac{\tau^2}{|b|^{4/3}}\left[1 + O(\tau^2)\right].
\label{h_4sol_tau}
\eeq

Much in the same way as in the large $r$ limit it is easy to see that the dilaton equation reduces to the equation 
\eqref{munu_initial} on the solution \eqref{dilaton_solution}.

Consider now Einstein's equations with internal indices. 
Let  index $a$ ($b$) denote differentials $e_{\psi},g_3,g_4$, while index $i$ ($j$) denote $g_1,g_2$.
Then we can calculate leading contributions to Christoffel symbols with $\tau$ indices at small $\tau$, namely
\beqn
&& \Gamma^{\tau}_{ij}= -g^{(c)}_{ij}\,\frac{g^{\tau\tau}_{c}\,h_2^{1/2}}{ah_1^{12}}\left(\frac1{\tau}+\frac14\,\frac{h_2'}{h_2}\right) , \qquad  
\Gamma^{\tau}_{ab}= -g^{(c)}_{ab}\,\frac{g^{\tau\tau}_{c}}{a}\,\frac14\,\frac{h_1'}{h_1}\,,
\nonumber\\
&&\Gamma^{j}_{\tau i}=\Gamma^{j}_{i \tau}=\delta^{j}_{i}\left(\frac1{\tau}+\frac14\,\frac{h_2'}{h_2}\right),
\qquad \Gamma^{b}_{\tau a}=\Gamma^{b}_{a \tau}= \delta^{b}_{a}\,\frac14\,\frac{h_1'}{h_1}\,,
\nonumber\\
&&\Gamma^{\tau}_{\tau\tau}=\frac12\,\frac{a'}{a}+\frac14\,\frac{h_1'}{h_1}, \qquad \Gamma^{n}_{\tau\tau}
= \Gamma^{\tau}_{\tau n}= \Gamma^{\tau}_{n\tau}=0,
\qquad n\neq r,
\label{Gammas_tau}
\eeqn

Then nonzero  components of the  Ricci tensor to  the leading order in $\tau$ take the form 
\beqn
&& R_{ij}\approx g^{(c)}_{ij}\,g^{\tau\tau}_{c}\left\{ \frac1{\tau^2}\,\left(\frac{ah_1^{1/2}}{h_2^{1/2}}-1\right) 
-\frac14 \Delta h_2 -\frac1{2\tau}\, h_2' -\frac1{2\tau}\, h_1'+\frac1{\tau}\, h_4' +\frac1{2\tau}\,a'
\right\},
\nonumber\\
&& R_{ab}\approx g^{(c)}_{ab}\,\,g^{\tau\tau}_{c}\left\{  -\frac14 \Delta h_1  \right\}, 
\nonumber\\
&& R_{\tau\tau}\approx   -\frac12 \Delta h_2 +\frac14 \Delta h_1 - h_1'' + h_4'' +\frac1{\tau}\,a' \, .
\label{Ricci_internal_tau}
\eeqn
 Here again we used that Ricci tensor is zero if all warp factors are equal to unity and the dependence on $h_4$ is found using formulas in \cite{deWit}. 

For the first solution in \eqref{1st_sol_tau} r.h.s.'s of Einstein's equations \eqref{Einstein} take the form
\beqn
&& R_{ij}= \frac1{ 2^4\,3^2}\,g^{(c)}_{ij} e^{-\Phi_0}\, (H_3^{(1)})^2,
\nonumber\\
&& R_{ab}= \frac{5}{2^4\,3^2}\,g^{(c)}_{ab} \,e^{-\Phi_0}\, (H_3^{(1)})^2 
\nonumber\\
&&R_{\tau\tau}= \frac1{ 2^4\,3^2}\,g^{(c)}_{\tau\tau} e^{-\Phi_0}\, (H_3^{(1)})^2,
\label{Einstein_eq_tau1}
\eeqn
where $(H_3^{(1)})^2$ is given by \eqref{H_3^2_tau} with $\mu_2=0$.

Solutions to these equations are given by
\beqn
&& h_1^{(1)}=1- \frac{5}{2^{2/3}\,3^{1/3}}\,\frac{\mu_1^2}{g_s}\,\frac{\tau^2}{|b|^{4/3}} + \cdots
\nonumber\\
&& h_2^{(1)}=1- \frac{5}{13}\,\frac{3^{2/3}}{2^{2/3}}\,\frac{\mu_1^2}{g_s}\,\frac{\tau^2}{|b|^{4/3}} + \cdots
\nonumber\\
&& a^{(1)}= 1 + \frac{5}{13}\,\frac{2^{1/3}}{3^{1/3}}\,\frac{\mu_1^2}{g_s}\,\frac{\tau^2}{|b|^{4/3}} + \cdots \,,
\label{warp_sol_tau_1}
\eeqn
where dots stand for corrections in powers of $\tau$ and powers of $\mu^2/|b|^{4/3}$.

For the second solution \eqref{2nd_sol_tau} r.h.s.'s of Einstein's equations \eqref{Einstein} has the form
\beqn
&& R_{ij}= \frac{5}{ 2^4\,3^2}\,g^{(c)}_{ij} e^{-\Phi_0}\, (H_3^{(2)})^2,
\nonumber\\
&& R_{ab}= \frac{1}{2^4\,3^2}\,g^{(c)}_{ab} \,e^{-\Phi_0}\, (H_3^{(2)})^2 
\nonumber\\
&&R_{\tau\tau}= \frac{5}{ 2^4\,3^2}\,g^{(c)}_{\tau\tau} e^{-\Phi_0}\, (H_3^{(2)})^2,
\label{Einstein_eq_tau2}
\eeqn
where $(H_3^{(2)})^2$ is given by \eqref{H_3^2_tau} with $\mu_1=0$.

For this case solutions take the form
\beqn
&& h_1^{(2)}=1- \frac{1}{2^{2/3}\,3^{1/3}}\,\frac{\mu_2^2}{g_s}\,\frac{\tau^2}{|b|^{4/3}} + \cdots
\nonumber\\
&& h_2^{(2)}=1- \frac{3^{2/3}}{2^{2/3}}\,\frac{\mu_2^2}{g_s}\,\frac{\tau^2}{|b|^{4/3}} + \cdots
\nonumber\\
&& a^{(2)}= 1 + \frac{2^{1/3}}{3^{1/3}}\,\frac{\mu_2^2}{g_s}\,\frac{\tau^2}{|b|^{4/3}} + \cdots \,.
\label{warp_sol_tau_2}
\eeqn

Note that much in the same way as for the large $r$ case the first and the third Einstein's equations in 
\eqref{Einstein_eq_tau1} and \eqref{Einstein_eq_tau2} coincides and give rise to conditions for the same combination 
$(3c_2 -2\tilde{A})$, where $c_2$ and $\tilde{A}$ are coefficients in front of $\tau^2$ for $h_2$ and $a$. As we explained 
before this is due to the possibility of redefinition of the radial coordinate ($\tau$ in the present case), see 
Sec. \ref{sec_warp_r}. To fix the definition of $\tau$ we require that the orthogonal combination to the one above  
is zero, $(2c_2 +3\tilde{A})=0$. This gives warp factors $h_2$ and $a$ presented in \eqref{warp_sol_tau_1} and 
\eqref{warp_sol_tau_1}.

We see that warp factors in \eqref{warp_sol_tau_1} and \eqref{warp_sol_tau_2} as well as $h_4$ \eqref{h_4sol_tau} and the dilaton \eqref{dilaton_solution} have smooth behavior at small $\tau$ and do not develop singularities provided  $\mu^2\ll |b|^{4/3}$. They can be found order by order at $\mu^2\ll |b|^{4/3}$ using perturbation theory in gravity equations.

\subsection{The scalar potential  at large $|b|$}

To find the scalar potential for the complex structure modulus $b$ we substitute solutions found above in this section
into Eq. \eqref{pot_gen}. At the leading order in $\mu^2/|b|^{4/3}$ we can neglect warp factors and use metric of the deformed conifold together with leading order expression \eqref{H_3^2_tau}. Using \eqref{det} at small $\tau$ we get
\beq
V(b) = {\rm const}\, (\mu_1^2+\mu_2^2)\, \frac{T^2}{g_s^3}\, \tau_{{\rm max}}^3,
\label{tau3}
\eeq
where $\tau_{{\rm max}}$ is the infrared cutoff with respect to the radial coordinate $\tau$ related to $R_{\rm IR}$ as follows
\beq
|b|\cosh(\tau_{{\rm max}}) = \left(\frac23\right)^{\frac32}\,R_{\rm IR}^3,
\eeq
see \eqref{tau}. This potential was obtained in \cite{Y_NSflux} for the first solution for $H_3$ proportional to $\mu_1$.

As we already explained,  we expect that in our runaway vacuum $b$ is large, close to $R_{\rm IR}$, therefore
$\tau_{{\rm max}}$ is small. Expanding $\cosh{\tau}$ at small $\tau$ we get
\beq
\tau_{{\rm max}} \sim \sqrt{\frac{\left(\frac23\right)^{\frac32}R_{\rm IR}^3- |b|}{|b|}}.
\eeq
This gives  the potential for the baryon $b$ at large $|b|$
\beq
V(b) = {\rm const}\, \left(\mu_1^2 +\mu_2^2\right)\, \frac{T^2}{g_s^3}\, \left[\frac{\left(\frac23\right)^{\frac32}R_{\rm IR}^3
- |b|}{|b|}\right]^{\frac32}.
\label{potlargeb}
\eeq
We see that to minimize the potential above $|b|$ becomes large and approaches the infrared cutoff,
\beq
\bra|b|\ket = \left(\frac23\right)^{\frac32}\,R_{\rm IR}^3 \to \infty.
\label{VEVb}
\eeq
As we expected earlier in Sec.~\ref{sec:pot}, we get  a runaway vacuum. 

The corrections to the potential \eqref{potlargeb} arise from taking into account higher powers of $\tau$ in the deformed conifold metric as well as from warp factors and go in powers of $\tau_{{\rm max}}$ and in powers of $\mu^2/|b|^{4/3}$ respectively. Both type of corrections disappear at the runaway vacuum \eqref{VEVb}.

In fact, $\tau_{{\rm max}}^3$
which enters \eqref{tau3} is the volume of the  three dimensional cone bounded by the  sphere $S_2$ of the conifold with the maximum radius $\tau_{{\rm max}}$. It shrinks to zero as $b$ tends to its VEV \eqref{VEVb}.
To avoid singularities we can regularize the size of $S_2$  introducing small non-zero $\beta$, which  makes the conifold ''slightly resolved'', see \eqref{D-term}. We  take the limit $\beta\to 0$ at the last step. Then the value of the potential and all its derivatives vanish in the vacuum \eqref{VEVb} at $|b|= \bra|b|\ket$, for example 
\beq
V(b)|_{|b|= \bra|b|\ket} = {\rm const}\,  \left(\mu_1^2 +\mu_2^2\right)\, \frac{T^2}{g_s^3}\,  \frac{\beta^3}{R_{\rm IR}^{9/2}} \to 0.
\label{VatVEV}
\eeq
In particular, the mass term for $b$ is zero.  

Absence of warp factors and vanishing of the potential $V(b)$ together with all its derivatives at the runaway vacuum confirms that \ntwo supersymmetry is not broken in 4D SQCD.

To summarize, the $H_3$-form flux produces following effects.\\
(i) The  Higgs branch of the baryon $b$ in 4D SQCD is lifted. \\
(ii) The vacuum is of the  runaway type $\bra|b|\ket \to\infty$.\\
(ii) At the runaway vacuum warp factors tend to unity and the geometry becomes that of the deformed conifold.\\
(iii) At the runaway vacuum  the radial coordinate $\tau$ and the sphere $S_2$  of the conifold degenerates, while the radius of the sphere $S_3$ tends to infinity.

 We will interpret this degeneration in terms of \ntwo SQCD in the next section. 

\section {Interpretation in terms of 4D SQCD }
\label{sec:quarkmasses}
\setcounter{equation}{0}

\subsection{3-form flux in terms of quark masses}

As we already mentioned in the Introduction $H_3$-form flux was interpreted in terms of 4D SQCD in \cite{Y_NSflux} as switching on quark masses. The motivation is that the only scalar potential deformation allowed in 4D SQCD by \ntwo supersymmetry is the mass term for quarks. Field theory
arguments were used in \cite{Y_NSflux} to find a particular choice of nonzero quark masses associated with $H_3$.
In this section we briefly review this  interpretation.  For $N_f=4$ we have four complex mass parameters. However, a shift of the complex scalar $a$, a superpartner of  the U(1) gauge field, produces an overall shift of  quark masses.  Thus, in fact we have three independent complex mass parameters in our 4D SQCD. For example, we can choose  three mass differences 
\beq
m_1-m_2, \qquad m_3-m_4, \qquad m_1-m_3
\label{massdifferences}
\eeq 
as independent parameters.

On the string theory side our solution \eqref{H_3sol} for the 3-form $H_3$ is parametrized by two real parameters $\mu_1$ and 
$\mu_2$.
Thus, we expect that non-zero $H_3$-flux can be interpreted in terms of a particular choice of quark masses, subject to  two complex constraints.

One constraint follows from \eqref{m_b}. We have seen in the previous subsection that $H_3$ does not produces a mass term for the  $b$-baryon. This ensures that
\beq
m_1 +m_2 -m_3 -m_4 =0.
\label{constraint1}
\eeq

Another constraint  is
\beq
m_1 m_2-m_3 m_4 =0.
\label{constraint2}
\eeq
It is imposed to avoid infinite VEV of $\sigma$ (a scalar superpartner of the U(1) gauge field), 
 which would costs an infinite energy in the world sheet  \wcpt model at large $b$, see \cite{Y_NSflux} for details.

Solving two constraints above leads to two options for the choice of the quark masses
\beq
m_3=m_1, \qquad m_4=m_2
\label{option1}
\eeq
and 
\beq
m_3=m_2, \qquad m_4=m_1.
\label{option2}
\eeq
These two options are essentially the same, up to permutation of quarks $q^3$ and $q^4$. Let us choose the first option in 
\eqref{option1}.

The  arguments above lead to the conclusion that the $H_3$-flux can be interpreted in terms of the  single mass difference
$(m_1-m_2)$. We define a complex parameter  $\mu $ and  identify \cite{Y_NSflux}
\beq 
\mu \equiv \mu_1 + i\mu_2 = {\rm const}\,\sqrt{\frac{g_s^3}{T}}\, (m_1-m_2), \qquad m_3=m_1, \qquad m_4=m_2.
\label{mum}
\eeq
The potential  \eqref{pot_r} calculated at large $r$, $r\gg |b|^{1/3}$ takes the form 
\beq
V(b)=  {\rm const}\,T\,|m_1-m_2|^2\,\log{\frac{R^3_{IR}}{|b|}}.
\label{pot_r_m}
\eeq
Similar substitution can be done for the large-$b$ potential 
\eqref{potlargeb}.

\subsection{Degeneration of the conifold and flow to SQED}
\label{sec:degeneration}

Since our solution to the gravity equations is valid at 
\beq
\frac{|\mu|^2}{|b|^{4/3}} \sim \frac{|m_1-m_2|^2}{T\,|b|^{4/3}} \ll 1
\label{reg_valid_m}
\eeq
and VEV of $b$ goes to infinity ( see \eqref{VEVb}) we can use our solution at  arbitrary large fixed values of $(m_1-m_2)$.
In particular, if we take   $|m_1-m_2|\gg \sqrt{\xi}$ in 4D SQCD keeping the constraint \eqref{option1}  non-Abelian degrees of freedom decouple and U(2) gauge theory
flows to \ntwo supersymmetric QED with the gauge group U(1) and $N_f=2$ quark flavors. 
Off-diagonal gauge fields together with two quark flavors acquire large masses $\sim |m_1-m_2|$ \footnote{In addition to masses $m_G\sim g\sqrt{\xi}$ due to the Higgs mechanism, see \cite{SYrev} for a review.}  and decouple. 

What happens to the non-Abelian vortex string upon this decoupling? The string survives, but transforms into an Abelian 
string. To see this note, that if we say,  increase masses $m_2=m_4$ keeping $m_1=m_3=0$ fields $n^2$ and $\rho^4$ decouple in the world sheet \wcpt
model on the string and it flows into \wcpo model. The $D$-term condition \eqref{D-term} now reads
\beq
|n^1|^2 - |\rho^3|^2 =  {\rm Re}\,\beta \,.
\label{D-term_11}
\eeq
The number of real degrees of freedom in \wcpo model  is $4-1-1=2$ where 4 is the number of real degrees of freedom of $n^1$ and $\rho^3$ and we subtract 2 due to the $D$-term constraint  \eqref{D-term_11} and 
the U(1) phase eaten by the Higgs mechanism.

Physically \wcpo model describes an  Abelian semilocal vortex string supported in \ntwo supersymmetric U(1) gauge theory with 
$N_f=2$ quark flavors.  This vortex has no orientational moduli, but it  has one complex size modulus $\rho^3$, see 
\cite{AchVas,SYsem,Jsem}.  Thus, we see that upon switching on $(m_1-m_2)$ a non-Abelian string flows to an Abelian one.

The low energy \wcpo model is also conformal.  Moreover,  it was shown in \cite{AharSeib} that in the non-linear sigma model formulation it flows to a free theory on $\mathbb{R}^2$ in the infrared.
Thus, in fact, switching on $(m_1-m_2)$ with constraint  \eqref{option1} does not break the conformal invariance on the world sheet. It just reduces the number of degrees of freedom  transforming a non-Abelian string into an Abelian one. The string theory which one would  associate with the \wcpo model  is non-critical. 

The field theory physics described above supports our interpretation of the  $H_3$-form flux  on the conifold in terms of quark masses.
On the string theory side switching on $(m_1-m_2)$ is reflected in the degeneration of the conifold, which effectively reduces its dimension. Also in the limit 
$|b|\to\infty$ the radius of the sphere $S_3$ of the conifold  becomes infinite and it tends to a flat three dimensional space. This matches the field theory result \cite{AharSeib} that \wcpo model flows to a free theory in the infrared. It would be tempting to interpret the extra coordinate of the sphere $S_3$ of the conifold in the limit $|b|\to\infty$ as a Liouville coordinate for a non-critical string associated with the  \wcpo model. This is left for a future work.

We also note that Eq. \eqref{Sb} suggests that massless stringy baryon $b$ acquires infinitely strong interactions at the runaway vacuum \eqref{VEVb} and the associated physics is no longer under analytic control.

\section {Conclusions}
\label{sec:conclusions}
\setcounter{equation}{0}

In this paper we considered a deformation of the string theory for the critical non-Abelian vortex  supported in \ntwo SQCD with gauge group U(2) and $N_f=4$ quark flavors with  NS 3-form flux building on the results of our previous paper 
\cite{Y_NSflux}. Using supergravity approach we found a solution for the 3-form $H_3$ and its back reaction on the conifold metric and the dilaton at the first non-trivial order in the parameter 
$\mu^2/|b|^{4/3}$. The non-zero 3-form $H_3$ generates a potential  for the complex structure modulus $b$ of the conifold, which is interpreted as a massless BPS baryonic hypermultiplet in 4D SQCD at strong coupling.
This potential lifts the Higgs branch formed by VEVs of $b$ and leads to a runaway vacuum for $b$, $\bra |b|\ket\to\infty$.
The warp factors disappear at this runaway vacuum.

Following \cite{Y_NSflux} we interpret the 3-form $H_3$ as a quark mass deformation of 4D SQCD.    Field theory arguments
are used  to relate the 
3-form $H_3$ to the quark mass difference $(m_1-m_2)$, subject to the constraint \eqref{option1}, see \eqref{mum}.

At the run away vacuum the conifold degenerates to  lower dimensions. This qualitatively matches with a flow  to the \wcpo model  on the string world sheet, expected if one  switches on the mass difference   $(m_1-m_2)$ and decouples one $n$-field and one $\rho$-field. In 4D SQCD this  corresponds to a flow to \ntwo supersymmetric  QED with two charged flavors.

\section*{Acknowledgments}

The author is grateful to P. Gavrylenko, A. Gorsky,  E. Ievlev, A. Marshakov  and M. Shifman  for very useful and 
stimulating discussions. This work  was  supported by Skolkovo Institute of Science and Technology and  by the Foundation for the Advancement of Theoretical Physics and Mathematics ''BASIS'',  Grant No. 22-1-1-16-1.

\renewcommand{\theequation}{A.\arabic{equation}}
\setcounter{equation}{0}

\end{document}